\documentclass{elsart}
\usepackage{graphicx, amssymb}
\journal{Nuclear Instruments and Methods B}
\begin{document}
\begin{frontmatter}

\title{Free ion yield observed in liquid isooctane irradiated by $\gamma$ rays. Comparison with the Onsager theory}

\author[facul]{J. Pardo \corauthref{cor}},
\ead{juanpm@usc.es}
\author[facul]{F. G\'omez },
\corauth[cor]{Corresponding author.}
\author[facul]{A. Iglesias},
\author[hospi]{R. Lobato},
\author[hospi]{J. Mosquera},
\author[facul]{J. Pena},
\author[facul]{A. Pazos},
\author[hospi]{M. Pombar},
\author[facul]{A. Rodr\'{\i}guez}, 
\author[hospi]{J. Send\'{o}n}

\address[facul]{Universidade de Santiago,
Departamento de F\'{\i}sica de Part\'{\i}culas}

\address[hospi]{Hospital Cl\'{\i}nico Universitario de Santiago}
\thanks{Present address: Departamento de F\'{\i}sica de Part\'{\i}culas, Facultade de F\'{\i}sica, campus sur s/n, 15782 Santiago de Compostela (Spain).
 This work was supported by project PGIDT01INN20601PR from Xunta de Galicia}

\begin{abstract}
We have analyzed data on the free ion yield observed in liquid isooctane
 irradiated by $^{60}$Co $\gamma$ rays within the framework of the Onsager
 theory about initial recombination. Several distribution functions describing
 the electron thermalization distance have been used and compared with the
 experimental results: a delta function, a Gaussian type function and an
 exponential function.

A linear dependence between free ion yield and external electric field has
 been found at low electric field values ($E<1.2\;MV/m$) in excellent agreement
 with the Onsager theory. At higher electric field values we obtain a solution
 in power series of the external field using the Onsager theory.
\end{abstract}

\begin{keyword}
free ion yield, Onsager theory, isooctane, liquid-filled ionization chamber
\PACS 29.40.Ym \sep 72.20.Jv \sep 82.50.Gw
\end{keyword}
\end{frontmatter}

\section{Introduction}

Liquid filled ionization chambers are currently used in radiotherapy both for
 dosimetry (\cite{Wickman1}, \cite{Wickman2}, \cite{LA48}, \cite{Boellaard}) 
 and portal imaging \cite{vHerk}. One of the most commonly used liquids is
 isooctane (2,2,4 trimethylpentane). This nonpolar liquid has a quite constant stopping
 power ratio to water in a very wide energy spectrum (less than 3\% variation
 from 0.1 MeV to 20 MeV \cite{Wickman1}) and also its intrinsic mass density 
allows to achieve a spatial resolution in the millimeter range for therapy beams \cite{vHerk}.

Free ion yield $G_{fi}(E,T)$, is defined as the number of electron-ion pairs \linebreak
 escaping initial recombination per 100 eV of absorbed energy, and 
experimentally it is obtained from ionization current measurement. The
 knowledge of how it varies with temperature $T$, with external electric
 field $E$, and with radiation type, constitutes a fundamental problem to
 understand the operation of these devices. These dependences have been
 measured in a large number of liquids (\cite{holroyd}, \cite{onsager2}).
 The Onsager theory \cite{onsager} describes $G_{fi}(E,T)$, and has been
 tested in several liquids with good results (see for example \cite{onsager2}). The Onsager theory predictes a linear relationship between ionization current and electric field at low electric field values. The previous dependence 
can be obtained from numerical resolution of the Onsager theory.
 This linear behavior has to be extrapolated to very
 low electric field strength region because volume recombination depletes
 free charge density produced by radiation in the liquid.

In the current work we describe a detailed method to apply and to test the
 Onsager theory at the low field strength region and to obtain a precise
 dependence of $G_{fi}$ with electric field in the linear region for liquid
 isooctane.

\section{Theoretical considerations}

\subsection{Onsager theory of initial recombination}

When ionizing radiation interacts with a liquid, electrons released from \linebreak
 molecules  thermalize at a distance $r$, where electron and positive
 ion are still bounded by the Coulomb interaction. This will cause the
 recombination of the primary ionization pairs produced, which is called 
initial recombination. These effects are much more relevant in liquids than
 in gases due to the fact that mass density of liquid hydrocarbons is almost
 three orders of magnitude higher than density of gases at normal conditions.

Onsager solved the problem of the Brownian movement of an electron under the
 influence of both the ion Coulomb attraction and an external electric field
 $E$ \cite{onsager}. For isolated ionizations, initial recombination escape
 probability of an electron-ion pair within the Onsager theory is

\[ \hspace{2cm}\Phi (r,E,\Theta,T) = exp \{ -\frac{E r}{E_{0} r_{0}}(1 + \cos{\Theta}) \} \]
\[ \hspace{2cm} \times  \int_{r_{0}/r}^{\infty}  J_{0}\hspace{0.1cm} [\hspace{0.1cm} 2 \hspace{0.1cm} \{ -\frac{E r}{E_{0}r_{0}}(1 + \cos{\Theta})\hspace{0.1cm} s\hspace{0.1cm} \} ^{1/2} ]\]
\begin{equation}
 \hspace{4cm} \times \; exp(-s)\; ds
\label{onsager1}
\end{equation}

where $r$ is the initial separation between electron and ion (i.e.
 the thermalization distance), $\Theta$ is the angle between the line that
 initially connects the electron-ion pair and the external electric field.
 The variables $r_{0}=e^{2}/4\pi \epsilon \kappa T$ and
 $E_{0}=2\kappa T/er_{0}$ are the Onsager radius (the distance at which
 Coulomb energy equals thermal energy $\kappa T$) and the Onsager field (the
 field that would produce a voltage $2\kappa T/e$ over a distance $r_{0}$).
 Here $\epsilon$ is the liquid dielectric constant
 ($\epsilon=1.94 \cdot\epsilon _{0}$ for liquid isooctane at room
 temperature), $T$ is its temperature and $\kappa$ is the Boltzmann constant.
 Finally, $J_{0}$ denotes the zeroth-order Bessel function.

Mozumder \cite{mozumder} converted the integral of equation (\ref{onsager1})
 in an infinite series using properties of the Bessel functions. He also
 eliminated the angular dependence \linebreak
 averaging over a uniform distribution of $\cos\Theta$. Then, the angle
 averaged escape probability takes the following form,

\begin{equation}
\hspace{2cm}\Pi(r, E,T) = 1 - \frac{E_{0}r_{0}}{2Er} \sum_{n=0}^{\infty} A_{n}( \frac{2Er}{E_{0}r_{0}})\hspace{0.1cm} A_{n}(r_{0}/r)
\label{onsager2}
\end{equation}

 where $A_{n}(x)$ is the $n$ order incomplete $gamma$ function, which is given
 by:

\[\hspace{1.2cm} A_{n}(x)=\exp(-x)\sum_{k=2n+1}^{\infty} \frac{x^{k-n}}{(k-n)!}= \exp(-x)\sum_{m=n+1}^{\infty} \frac{x^{m}}{m!}\] 
\begin{equation}
\hspace{2.2cm}=1-\exp(-x) \lbrack \;1+ x + \frac{x^{2}}{2!}+ \cdots + \frac{x^{n}}{n!} \;\rbrack     
\label{onsager3}
\end{equation}

The next expression is more practical for numerical computation
 of $\Pi(r, E,T)$:

\begin{eqnarray}
\hspace{2cm}A_{n+1}(x) - A_{n}(x)& =& -(x^{n+1}/(n+1)!)\hspace{0.1cm} exp(-x)\\
A_{0}(x)&=&1-exp(-x) 
\label{onsager4}
\end{eqnarray}

Equation (\ref{onsager2}) is the most adequate formula for calculating escape
 probabilities for arbitrary values of initial separation and external
 electric field. In fact, it will be the formula that we will use for
 numerical calculations of escape probabilities. 
The expansion in power series of the external field is implicit in equation
 \ref{onsager2}. If we operate in equation (\ref{onsager2}) then we can
 obtain it explicitly:

\begin{equation}
\hspace{2cm}\Pi(r, E,T)= \exp(-r_{0}/r)\left[ 1+\sum_{n=1}^{\infty} \left(\frac{E}{E_{0}}\right)^{n}\hspace{0.05cm} B_{n}(r/r_{0})\right]
\label{onsager5}
\end{equation}

where $B_{n}(x)$ is a polynomial of order $n-1$ in $x$, which takes the
 following form:

\begin{equation}
\hspace{3cm}B_{n}(x)=\sum_{m=1}^{n}\left[\left(\sum_{k=m}^{n} F\hspace{0.05cm}_{k}^{n}\right) \frac{x^{(n-m)}}{m!}\right]
\label{Bn}
\end{equation}

The numerical coefficients $F_{k}^{n}$ were calculated by Mozumder
 \cite{mozumder}, and they are given by

\begin{eqnarray}
\hspace{4cm}&F_{k}^{n}& = 0  \hspace{0.5cm} for \hspace{0.2cm} k>n\\
& F_{n}^{n}&=\frac{2^{n}}{(n+1)!}
\end{eqnarray}
and
\begin{equation}
\hspace{3cm} F_{k-1}^{n}=F_{k}^{n}+ \frac{(-1)^{n-k+1}2^{n}}{k!(n-k+1)!}\hspace{0.5cm}for \hspace{0.2cm} k\le n
\end{equation}

We must keep in mind that thermalization distance is not the same for all
 electrons. Due to this fact a distribution function $f(r)$, such as to
 $\int_{0}^{\infty}f(r)dr=1$, is usually introduced $ad$ $hoc$ to describe
 electron thermalization distances. Several distribution functions had been
 used by different authors in the literature (see for example  \cite{onsager2}
 and \cite{mozumder}). In the current article we will test the three most used:

\begin{itemize}

           \item  delta function  \begin{equation}
 f(r,\rho)=\delta(r-\rho)\label{del} \label{delta} \end{equation}

 \item Gaussian type function
\begin{equation}   f(r,\rho,\sigma)=\left\{ \begin{array}{ll}
N\exp\lbrack-(r-\rho)^{2}/\sigma ^{2}\rbrack  \;\;\; &  \mbox{if $r\geq$ 0}\\
 0 & \mbox{if $r < 0$} \end{array} \right.
\label{gauss}
 \end{equation}

           \item  exponential function \begin{equation}
  f(r,\rho) = \left\{ \begin{array}{ll}
             \frac{1}{\rho}\exp ( -r/\rho )  \;\;\; & \mbox{if $r\geq$ 0}\\
               0 & \mbox{if $r < 0$} \label{exp} \end{array} \right.  
\end{equation}

\end{itemize}

The first and second distributions are characterized by a single parameter
 $\rho$. The Gaussian type distribution, in addition, has a dispersion
 parameter $\sigma$. However, in order to obtain a single parameter distribution we will correlate both parameters. In advance we will take
 $\sigma=0.25\cdot \rho$ such as \cite{mozumder}. This choice provokes that
 the difference between the normalization factor $N$, and the normal Gaussian
 normalization factor $1/\sqrt{2\pi}\sigma$, be less than $0.1\%$.

Within this framework we can write the escape probability averaged over the
 thermalization distances as:

\begin{equation}
\hspace{3cm}P_{esc}(E,T)=\int_{0}^{\infty}\Pi(r,E,T)\;f(r)\;dr
\label{pesc}
\end{equation}

An interesting property of the Onsager series is that the first expansion
 term of the escape probability does not depends on the thermalization
 distance $r$.
From equations (\ref{onsager5}) and (\ref{Bn}) we can derive that when the electric field is sufficiently low to verify

\[\hspace{3.5cm} 3(\frac{E_{0}}{E})\gg \mid 1- 2(\frac{r}{r_{0}})\mid \]

the $n=1$ term is much higher than the $n=2$ term in equation (\ref{onsager5}) (and hence than terms $n>2$). Then we can truncate the series to first order and so the escape probability rises linearly with the electric field:

\begin{equation}
\hspace{3cm}\Pi(E,T)= \exp(-r_{0}/r)\left( 1+\frac{E}{E_{0}} +\cdots \right)
\label{onsager6}
\end{equation}

 In this case the intercept-to-slope ratio $E_{c}$ is predicted to be the Onsager field $E_{0}$:

\begin{equation}
\hspace{3.5cm}E_{c}=E_{0}=\frac{8\pi(\kappa T)^{2}}{e^{3}}
\label{corte}
\end{equation}

This result is a powerful test to compare experimental data with the Onsager
 theory because it does not depend on the distribution function used to
 describe the electron thermalization distance.

\begin{figure}
\begin{center}
\includegraphics*[width=9.5cm]{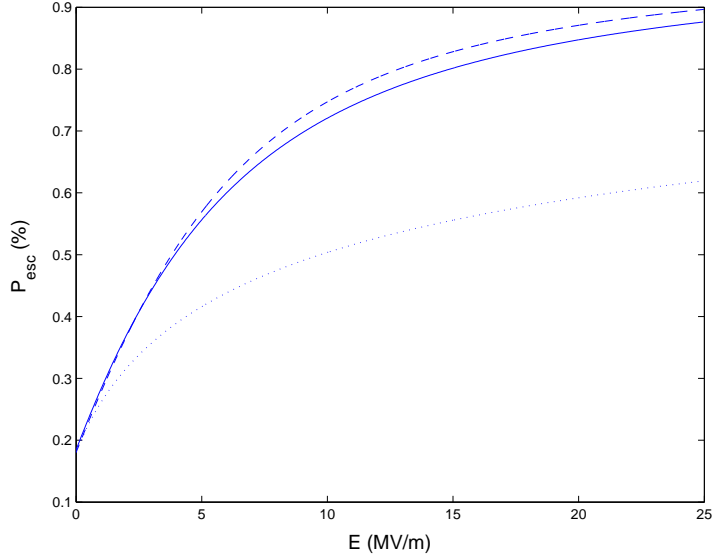}
\end{center}
\caption{\em Escape probability calculated using a Gaussian distribution
 (continuous line), an exponential distribution (dotted line) and a delta
 distribution (dashed line), plotted against electric field for liquid
 isooctane at $T=294$ K.}
\label{fig1}
\end{figure}

\begin{figure}
\begin{center}
\includegraphics*[width=9.5cm]{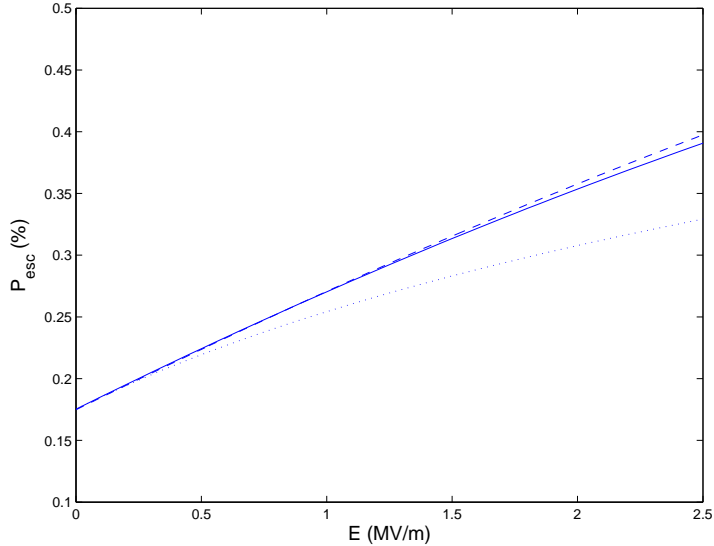}
\end{center}
\caption{\em. Escape probability calculated using a Gaussian distribution
 (continuous line), an exponential distribution (dotted line) and a delta
 distribution (dashed line), plotted against electric field for liquid
 isooctane at $T=294$ K. We can see how the relationship becomes linear at low
 electric fields (approximately $E<1.2\;MV/m$ for Gaussian type and
 delta distributions, $E<0.5\;MV/m$ for exponential distribution).}
\label{fig2}
\end{figure}

As we will see later, for isooctane at room temperature irradiated by $\gamma$
 photons, $r/r_{0}\sim 0.6$ and the linear approximation (\ref{onsager6}) is
 right at low electric fields.
 Figure \ref{fig1} shows escape probability variation with external electric
 field for the case of liquid isooctane at $T=294$ K. Figure \ref{fig2} shows the low electric field region where a linear relationship is expected.

\subsection{Free ion yield calculation within the Onsager framework} 
 Free ion yield $G_{fi}$, is defined as the number of electron-ion pairs
 created (i.e. those escaping initial recombination) in the ionization medium
 per 100 eV of absorbed energy. This magnitude plays a similar role to the
 $W$ factor in gases. However, $W$ is constant and free ion yield depends
 on temperature, on external electric field and on radiation type as escape
 probability does. Within the Onsager theory we can write 

\begin{equation}
\hspace{3cm}G_{fi}(E,T)=N_{tot}P_{esc}(E,T)
\label{gfi}
\end{equation}

where $N_{tot}$ is the total number of electron-ion pairs formed initially in
 the ionization medium (i.e. before initial recombination) per 100 eV of
 absorbed energy, and $P_{esc}$ is the escape probability described in the
 previous subsection.

Free ion yield at zero external electric field is denoted as $G_{fi}^{0}$, and
 takes the following form:

\begin{equation}
\hspace{3cm}G_{fi}^{0}= N_{tot}\int_{0}^{\infty}\; f(r)\; \exp(-r_{0}/r)
\label{freeion1}
\end{equation} 

If we are working at electric field values for which the relation between
 escape probability and electric field is linear, then we can write

\begin{equation}
\hspace{4cm}G_{fi}= G_{fi}^{0}\;+\;aE
\label{freeion2}
\end{equation}

and the intercept-to-slope ratio from equation (\ref{corte})

\begin{equation}
\hspace{4cm}E_{c}=\frac{G_{fi}^{0}}{a}=E_{0}
\label{corte2}
\end{equation}

\section{Experimental results}

\begin{figure}
\begin{center}
\includegraphics*[width=10cm]{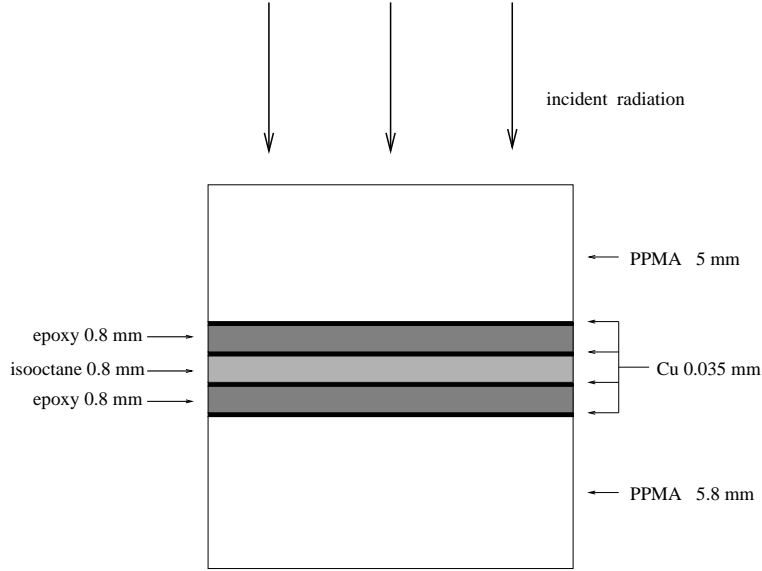}
\end{center}
\caption{\em Scheme of the liquid ionization chamber cross section.}
\label{PCB}
\end{figure}

In order to measure the ionization current from an isooctane layer under
 irradiation we built a square shaped parallel plate liquid ionization chamber.
 The chamber walls were fabricated using FR4 fiber glass reinforced epoxy
 copper clad on both sides, covering a total area of 4.08 cm$\times$4.72 cm.
 The FR4 thickness was 0.8 mm while the copper layer was 35 $\mu$m thick. The
 two chamber walls were glued on both sides of an epoxy plate spacer to
 provide the 0.8 mm isooctane gap. To guarantee a constant dose rate along
 the gap, the detector was inserted between two PMMA\footnote{Polimethylmetacrylate.} plates of 5 mm and 5.8 mm thickness, the first one on the incident
 beam side. As ionization medium we used Merk's liquid isooctane\footnote{Isooctane Merk Uvasol quality grade.} with an estimated purity of $99.8\%$.
Figure \ref{PCB} shows a scheme of the device cross section.
 
Experimental tests of the chamber were made in the $^{60}$Co unit of the
 Complexo Hospitalario Universitario de Santiago (CHUS). The radiation field
 was set to 10 cm$\times$10 cm at the isocenter\footnote{80 cm from the cobalt
 source.} to cover the whole chamber.
In order to obtain isooctane free ion yield  we measured the ionization
 current produced in the whole chamber using a nanoammeter Phillips Fluke PM
 2525, for several polarization voltages. High voltage was supplied by a
 CAEN N471A NIM module.

\begin{figure}
\begin{center}
\includegraphics*[width=10cm]{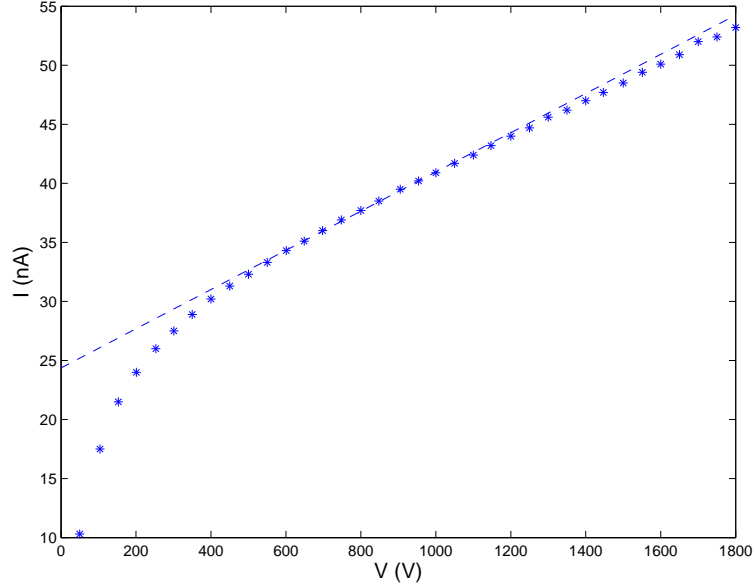}
\end{center}
\caption{\em Ionization current measured in the chamber against polarization
 voltage. The continuous line shows the linear extrapolation at low field
 strength region.}
\label{current}
\end{figure}

 Distance between the detector and the cobalt source was set to 130 cm
 (equivalent to a dose rate around 0.4 Gy/min). The dose rate was chosen to
 have a negligible volume recombination in the upper part of the ionization
 current vs. voltage curve. Figure \ref{current} shows the experimental data
 obtained.

If volume recombination and space charge effects can be ignored, the
 ionization current is proportional to the number of electron-ion pairs
 released in the liquid per unit time and unit volume $N_{in}$ (initial
 recombination is included in $N_{in}$):

\begin{equation}
\hspace{4.5cm}I=ehAN_{in}
\label{current1}
\end{equation}

where $e$ is the electron charge, $h$ is the isooctane gap and $A$ is the
 detector area. From this equation the free ion yield can be calculated as 
\begin{equation}
\hspace{5cm} G_{fi} = \frac{I}{e\;\Delta  \varepsilon} 
\label{gfi2}
\end{equation}

where $\Delta \varepsilon $ is the energy deposited in the medium per second.
 In this case  $\Delta \varepsilon =(4.79\pm0.19)\cdot 10^{11}$ 100 eV/s,
 that was calculated through the EGSnrc code. 

To apply equation (\ref{gfi2}) we require a charge collection efficiency
 higher than 99\% and also that field screening be negligible. To calculate
 this limit we used a numerical simulation of the charge carriers transport.
 When the distance between the cobalt unit and the detector is 130 cm, and the
 polarization voltage is higher than 600 V, general charge collection
 efficiency is higher than $99\%$. This agrees with the Greening theory
 (\cite{greening1}, \cite{greening2})  about general charge collection
 efficiency. Within this theory the polarization voltage that must be applied
 to obtain an efficiency $f$, is

\begin{equation}
\hspace{4.2cm}V^{2} = \frac{1}{6}\frac{m^{2}h^{4}eN_{in}}{(\frac{1}{f}-1)}\\
\label{eq_gr}
\end{equation}

with

\[\hspace{4.5cm}m^{2}=\frac{\alpha}{ek_{+}k_{-}}\]

In these equations $k_{+}$ and $k_{-}$ are the mobilities of the positive and
 negative charge carriers and $\alpha$ is the volume recombination constant
 (we used $k_{+}=k_{-}=3.2 \cdot 10^{-8}$ m$^{2}$V$^{-1}$s$^{-1}$ and
 $\alpha=5.9 \cdot 10^{-16}\;$m$^{3}$s$^{-1}$)\footnote{Values obtained under
 irradiation with X rays, in agreement with \cite{Wickman3}.}. Introducing
 the numerical data of the experimental set-up in equation \ref{eq_gr} we 
obtain $f\ge0.99$ for $V\ge$625V. Then we only apply equation (\ref{gfi2}) to data for $V\geq$600 V.

 For lower voltages the free ion yield has to be extrapolated because volume
 recombination and space charge effects provoke charge losses and field
 screening, and equations (\ref{current1}) and (\ref{gfi2}) no longer hold.
We expect from section 2.1 a linear relationship between the free ion yield
 and the electric field at low electric field values. Figure \ref{current}
 shows this linear relationship between 600 V and 1000 V. At higher voltages
 the relationship begins to deviate from linearity as shows the figure. Then,
 we use data in the range 600 V $\leq V\leq$1000 V to extrapolate the free
 ion yield at low electric field region.

 This experimental linear relationship  between free ion yield and external
 electric field $E$, is
\begin{equation}
\hspace{2cm}G_{fi}(E)=(0.32\pm 0.02)+(1.73\pm 0.09)\cdot 10^{-7}\cdot E
\label{gfi_exp}
\end{equation}

where $E$ is given in V/m and the free ion yield in pairs/100 eV. Equation
 (\ref{gfi_exp}) is valid up to $E=1.2$ MV/m with a confidence level of 96\%. For higher values
 be have to take into account more terms in the equation (\ref{onsager5}).  

For isooctane the total number of electron-ion pairs produced per 100 eV of
 absorbed energy is $N_{tot}=1.83$ (also calculated with the EGSnrc code).
 Inserting this value and the obtained $G_{fi}^{0}$ in equation (\ref{freeion1}
) we can obtain the parameter $\rho$, for the delta (\ref{delta}), Gaussian
 type (\ref{gauss}) and exponential (\ref{exp}) distribution functions. The
 results are $\rho=168\pm6$ \AA$\;$for first and second distributions, and
 $\rho=178\pm10$ \AA$\;$for the exponential one.

\begin{figure}{}
\begin{center}
\includegraphics*[width=10cm]{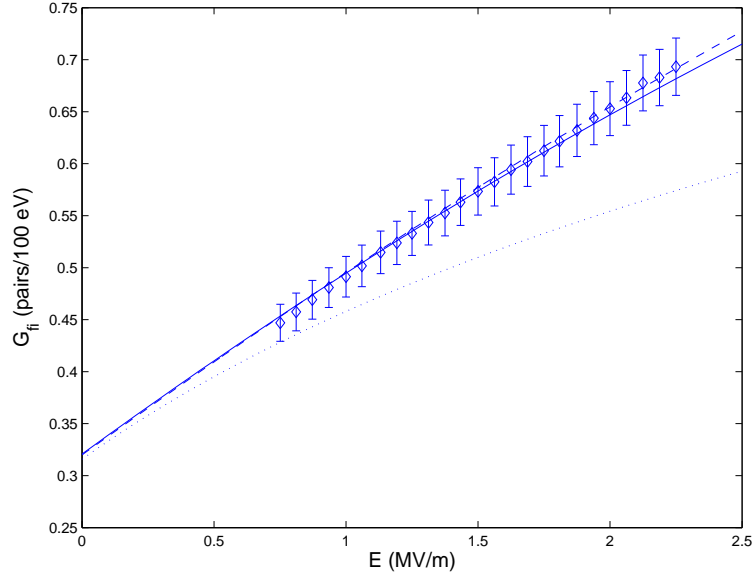}
\end{center}
\caption{\em Free ion yield against external electric field in the linear
 region. The diamond points show the experimental points for E$\geq$0.75 MV/m
 (for lower fields we have to extrapolate using equation (\ref{gfi_exp})). The dotted,
 dashed and continuous lines correspond to the theoretical prediction using
 exponential, delta and Gaussian type distribution functions respectively.}
\label{fig_gfi}
\end{figure}

Taking the numeric values for $N_{tot}$ and $\rho$ we can calculate the
 theoretical prediction for free ion yield within the Onsager theory using
 equations (\ref{gfi}), (\ref{pesc}) and (\ref{onsager2}), and compare these
 theoretical results with experimental data. Figure \ref{fig_gfi} shows
 results obtained using the three considered distribution functions. Delta
 and Gaussian type distributions  agree with the experimental data, but not
 the exponential distribution.

The intercept-to-slope ratio from equation (\ref{gfi_exp}) is  $E_{c}=(1.83\pm
 0.12)$ MV/m. Within the Onsager framework we obtain (see equation
 (\ref{corte}) and figure \ref{fig2}), $E_{c}=E_{0}=(1.74\pm 0.02)$ MV/m at
 the room temperature, $T=(294\pm 2)$ K. Experimental value and theoretical
 value for $E_{c}$ are clearly in agreement.

\section{Conclusions}

We have obtained and analyzed data on isooctane free ion yield irradiated by
 $\gamma$ photons, from a cobalt source, within the framework of the Onsager
 theory. Three distribution functions (describing separation distance between
 electron-ion pairs when thermalization is achieved) have been considered:
 a delta function, a Gaussian type function and an exponential function.
 The first and the second  describe data correctly in the covered electric
 field range, but not the exponential function.

This fact means that free ion yield depends in a fundamental way on the choice
 of the distribution function, which is not predicted by the theory. The good
 agreement between the experimental data and the theoretical prediction using
 a delta or a Gaussian type distribution with a dispersion parameter
 $\sigma =0.25\cdot\rho$ seems to show that in this case $f(r)$ is a Gaussian
 type function with a small dispersion parameter.

If electron would suffer a large number of independent collisions before
 thermalization, a Gaussian distribution function for the thermalization
 distances will agree with the central limit theorem. The lack of data about
 the nonpolar liquids cross sections makes difficult to obtain models
 describing the nature of $f(r)$. Some computer simulations have
 been made (see for example \cite{Musolf}) in this sense, however more
 theoretical and numerical work is needed in this area.

On the other hand, the theoretical prediction for the intercept-to-slope ratio
 $E_{c}$, is in agreement with the experimental value. This is an important
 result to check the Onsager theory because it does not depend on $f(r)$.
\section{Acknowledgments}

The authors express their gratitude to L.M. Varela from the Department of
 Condensed Matter Physics from the University of Santiago for his useful
 comments about theory of liquids.

\end{document}